\documentclass[12pt]{article}
\usepackage{amssymb}

\usepackage{epsfig}

\def\beq{\begin{equation}}
\def\eeq{\end{equation}}
\input{tcilatex}
\begin{document}

\title{\textbf{Generalized Borel Transform Technique in Quantum Mechanics \thanks{%
Partially supported by CONICET and ANPCyT-Argentina.}}}
\author{L.N. Epele$^a$, H. Fanchiotti$^a$, C.A. Garc\'{\i }a Canal$^a$, M. Marucho$%
^b $ \\
\\
$^a$Laboratorio de F\'{\i }sica Te\'{o}rica\\
Departamento de F\'{\i}sica \\
Universidad Nacional de La Plata \\
C.C. 67 - 1900 La Plata \\
Argentina\\
\\
\\
$^b$The Maurice Morton Institute of Polymer Science\\
The University of Akron\\
Akron, OH 44325-3909, USA\\
marucho@polymer.uakron.edu}
\date{}
\maketitle

\begin{abstract}
We present the Generalized Borel Transform (GBT). This new approach allows
one to obtain approximate solutions of Laplace/Mellin transform valid in
both, perturbative and non perturbative regimes. We compare the results
provided by the GBT for a solvable model of quantum mechanics with those
provided by standard techniques, as the conventional Borel sum, or its
modified versions. We found that our approach is very efficient for
obtaining both the low and the high energy behavior of the model.
\end{abstract}

\vspace{5mm} \noindent Classification: 11.15 Bt - 12.38 Cy\newline
 \bigskip

\newpage

Recently, Penin and Pivovarov \cite{rusos} presented a numerical analysis of
renormalon techniques in quantum mechanics. They used a simple solvable
model, namely a delta function scattering potential, to address the problem
of resummation of perturbative series \cite{hardy}. In fact, this quantum
mechanics potential can be considered as a confining one, mimicking the long
sought property of Quantum Chromodynamics (QCD). This is one of the many
examples where the finite order perturbation theory predictions present
uncertainties comparable with experimental errors \cite{error}. In all these
cases one is forced to take into account, in a way or another,
non-perturbative contributions. In connection with these difficulties, it is
certainly instructive to study exactly solvable models so that the
efficiency and precision of the different proposals \cite{rusos}\cite
{propuestas} can be quantitatively checked. This analysis can help us to
define criteria for selecting the appropriate approach to be used in more
realistic cases as QCD is.

In reference \cite{rusos} the delta function model is tackled by means of
standard Borel summation techniques combined with the renormalon approach 
\cite{beneke}. On the other hand, we have recently introduced \cite{nos} a
Generalized Borel Transform (GBT) that avoids the implementation of a
perturbative expansion. This proposal was successfully applied \cite{nosrich}
to obtain an analytic expression for the heavy quark-antiquark potential,
valid for all distances.

The purpose of this paper is to discuss the use and advantages of the GBT in
the case of quantum mechanics and in particular in connection with the model
studied in ref. \cite{rusos}. We are able to obtain analytic expressions
valid in all the range of the variables.

We start by briefly summarizing ref. \cite{rusos} in order to prepare the
basic material of the delta potential that is further treated with our GBT
approach.

The quantum mechanics model, analyzed first in the scattering region, is
defined through the singular potential

\begin{equation}
V\left( r\right) =v\delta \left( r-a\right) \quad ;\quad v,a=cte  \label{pot}
\end{equation}
and can be considered as a kind of confining (not completely) interaction.
Usually, the dispersion relation related to this kind of singular potential
is analyzed in analogy with dispersion relations in elementary particle
theory \cite{dispersion}. Here, to simplify the presentation, we consider
only the $s$ - wave amplitude and the study of the wave function at the
origin. To analyze the scattering of a wave packet, one considers an
integral of the form

\begin{equation}
\Psi \left( Q\right) \equiv 1+F\left( Q\right) =\int_0^\infty \psi \left(
q\right) W\left( q,Q\right) dq  \label{paquete}
\end{equation}
where,

\[
W\left( q,Q\right) \equiv Q\frac{\exp \left( -Q/q\right) }{q^2} 
\]
is a momentum distribution function normalized to one and,

\begin{equation}
\psi \left( q\right) =\left[ 1+\frac v{2q}\left( 1-\exp \left( -2qa\right)
\right) \right] ^{-1}  \label{plana}
\end{equation}
is the exact solution for the scattering of a plane wave in the "Euclidean''
region where the momentum $q>0$ is obtained through a Wick rotation. This is
possible whenever $m=1/a>\left| v\right| $ and consequently there are no
bound states present.

The perturbative solution in eq. (\ref{paquete}) suffers of the same
difficulties present in perturbative QCD and it mimics some general features
of renormalons.

We start the analysis with the perturbative treatment in order to make
explicit the difficulties of the standard method. At high energies $\left(
q\gg m\right) $ one can write

\begin{equation}
\psi ^{as}\left( q\right) =\sum_{n=0}^\infty \left( -\alpha \left( q \right)
\right) ^n  \label{suma}
\end{equation}
where $\alpha \left( q\right) \equiv v/2q$ is the natural parameter of
expansion in expression (\ref{plana}).

Replacing now the series (\ref{suma}) into (\ref{paquete}) one obtains

\begin{equation}
\Psi ^{as}\left( q\right) =\sum_{n=0}^\infty n!\left( -\alpha \left(
q\right) \right) ^n  \label{sumafi}
\end{equation}
For $v>0$ , the theory is Borel summable \cite{beneke} and the conventional
Borel transform is

\begin{equation}
B\left( s\right) =\sum_{n=0}^\infty \left( -s\right) ^n=\frac
1{s+1}\rightarrow \Psi _S^{as}\left( q\right) =\frac 1{\alpha \left(
q\right) }\int_0^\infty B\left( s\right) \exp \left( -s/\alpha \left(
q\right) \right)  \label{sumaborel}
\end{equation}
Consequently, the approximate solution (\ref{sumafi}) results

\begin{equation}
\Psi _S^{as}\left( q\right) =\frac 1{\alpha \left( q\right) }\exp \left(
1/\alpha \left( q\right) \right) \func{Ei}\left( 1,1/\alpha \left( q\right)
\right)  \label{fias}
\end{equation}
where \cite{NU} 
\begin{equation}
\func{Ei}\left( n,x\right) \equiv \int_1^\infty \frac{\exp \left( -tx\right)
dt}{t^n}\quad n=1,2..\quad x>0  \label{expint}
\end{equation}

A numerical comparison of the approximate expression (\ref{fias}) with the
exact solution (\ref{paquete}), presented in Figure 1, clearly shows that it
is not a good approximation for intermediate values of $q\sim m$.

An alternative approach \cite{grun} uses a modified perturbation theory and
provides results of better precision at very small momenta. This technique
starts by choosing, instead of $\alpha(q)$, a modified expansion parameter
of the form

\[
\alpha _{\mu} \left( q\right) \equiv \frac v{2q+\mu }\left( 1+\frac{\mu -v}{%
2q+\mu }\right) 
\]
where $\mu =\mu \left( m,v\right) $ is fitted using experimental
information. In this model, the exact solution $\Psi\left( Q\right)$ given
in eq. (\ref{paquete}), plays the role of experiment data.

The accuracy of the approximation and the optimal value of $\mu $ clearly
depend on the range where data have to be fitted. Unfortunately, the value
that optimize the result at very small $Q,$ does not fulfil this requirement
at very large $Q $ as it is evident in Figure 1.

Hence, the method and its modification work reasonably well in providing an
approximation to the exact solution but each one in different regimes of $Q$%
. They are not able to reproduce simultaneously, with sufficient precision,
the exact result in both IR and UV regimes.

We have found that the Generalized Borel Transform \cite{nos}\cite{nosrich}
copes with that difficulty. This GBT approach was designed to obtain the
approximate solution of any Laplace/Mellin transform in the wide range of
the parameter involved.

In fact, when using our proposal, one is performing a whole class of
transformations, depending on a parameter that we call $\lambda$.\ As it
should be, the result does not explicitly depends on $\lambda$.
Consequently, one can choose for this parameter the best adapted value in
each particular problem under consideration. In practice, as one is
performing an approximate calculation, one can get rid of $\lambda$ by means
of a suitable saddle-point like technique.

Let us briefly present our technology in connection with a Laplace-Mellin
transform presents when treating the quantum mechanical problem under
analysis. In this case, we typically face a transformation like 
\begin{equation}
S\left( g,a,n\right) =\int_0^\infty x^nH\left( x,a\right) \exp \left(
-gx\right) dx\qquad;\qquad g>0  \label{int}
\end{equation}
where we have explicitly extracted a factor $x^n$ from the function to be
transformed, because, at it will became clear after, this operation
facilitates the implementation of our proposal.

The Generalized Borel Transform (GBT) of $S$ is defined as

\begin{equation}  \label{gbt}
B_{\lambda} \left( s,a,n\right) \equiv \int\limits_0^\infty \exp \left[
s/\eta \right] \left[ \frac 1{\lambda \eta }+1\right] ^{-\lambda s}S\left(
g,a,n\right) d\left( 1/\eta \right) \quad;\quad \func{Re}\left( s\right) <0
\label{borel}
\end{equation}
$\lambda $ being any real positive non zero value and where we have defined 
\begin{equation}
1/\eta \equiv \lambda \left[ \exp \left( g/\lambda \right) -1\right]
\label{use}
\end{equation}
This particular election of $\eta$ allows one to define the function $%
u_{\lambda} \left( g\right) $ monotonically increasing in the interval $%
0<u_{\lambda} \left( g\right) <\infty $, namely

\begin{equation}  \label{ug}
u_{\lambda} \left( g\right) \equiv \frac{1}{\eta } -\lambda \ln \left[ \frac
1{\lambda \eta }+1\right]
\end{equation}
Consequently $B_{\lambda} (s)$ results in

\[
B_{\lambda} \left( s,a,n\right) =\int\limits_0^\infty \exp \left( su\right)
S\left[ g_{\lambda} \left( u\right) ,a,n\right] \left[ 1+\lambda\, \eta
\left( u\right) \right] du\quad;\quad \func{Re}\left( s\right) <0 
\]
The last expression can be written as a Laplace transform 
\[
B_{\lambda} \left( s,a,n\right) =\int\limits_0^\infty \exp \left( su\right)
L_{\lambda} \left[ g\left( u\right) ,a,n\right]\, du\quad; \quad \func{Re}%
\left( s\right) <0 
\]
of the function $L_{\lambda} \left[ g\left( u\right) ,a,n\right] $ there
implicitly defined.

Replacing (\ref{int}) and (\ref{use}) in equation (\ref{borel}) we can
easily test that the GBT is an analytic function on the negative Borel
half-plane, such that its extension to the other half-plane also exists and
is also analytic with a cut on the real positive axis. From this
observation, $S\left( g,a,n\right) $ can be expressed in terms of the
inverse Laplace transform integrated on the above mentioned cut \cite
{nosrich}

\begin{equation}
S\left( g,a,n\right) =\frac 1{\left[ \lambda \eta +1\right]
}\int\limits_0^\infty \exp \left[ -su_\lambda \left( g\right) \right] \Delta
B_\lambda \left( s,a,n\right) ds  \label{fin}
\end{equation}

The main advantages of this proposal comes from the fact that it allows one
to perform the calculations in terms of the parameter $\lambda$. Moreover,
as it will become clear below, this approach avoids the implementation of a
perturbative expansion. Each value of the parameter $\lambda $ defines a
particular Borel Transform. This can be summarize by writing $S\left(
g,a,n\right) =T_{\lambda} ^{-1}\left[ T_{\lambda} \left( S\left(
g,a,n\right) \right) \right] $ where $T_{\lambda} \left( S\left(
g,a,n\right) \right) \equiv B_{\lambda} \left( s,a,n\right) $ .

The discontinuity of the $B_{\lambda} \left( s,a,n\right) $ can be expressed
as

\[
\Delta B_{\lambda} \left( s,a,n\right) =2\lambda \int\limits_{-\infty
}^\infty dw\exp \left[ T\left( w,\lambda ,s,a,n\right) \right] 
\]
where

\begin{eqnarray*}
T\left( w,\lambda ,s,a,n\right) &\equiv & -\ln \left\{ \Gamma \left[ \lambda
\left( s+x\left( w\right) \right) \right] \right\} +\left\{ \lambda \left[
s+x\left( w\right) \right] -1\right\} \ln \left( \lambda s\right) -\lambda s
\\
&& +w+\ln \left[ x^nH\left( x,a\right) \right]
\end{eqnarray*}
with $x=\exp \left( w\right)$ and $\Gamma$ represents de Euler gamma
function. After the change of variables $s=\lambda \exp \left( t\right)$ one
gets

\begin{equation}
S\left( g,a,n\right) =2\lambda ^2\left( 1-\exp \left( -g/\lambda \right)
\right) \int\limits_{-\infty }^\infty \int\limits_{-\infty }^\infty \exp
\left[ G\left( w,t,g,\lambda ,a,n\right) \right] dwdt  \label{doble}
\end{equation}
with

\begin{equation}
G\left( w,t,g,\lambda ,a,n\right) =-s\left( t\right) u_{\lambda} \left(
g\right) +t+T\left[ w,\lambda ,s\left( t\right) ,a,n\right]  \label{argexp}
\end{equation}

In this way we can obtain the dominant contribution of the double integral
using the steepest descent technique in the combined variables $\left[
t,w\right] $. In so doing, one first computes the saddle point $t_o\left(
g,a,n\right) $ and $w_o\left( g,a,n\right) $ in the limit $\lambda \gg 1$.
In this case, the saddle point is 
\begin{equation}
t_o\left( g,a,n\right) =\ln \left\{ \frac{x_o^2\left( g,a,n\right) }{f\left[
x_o\left( g,a,n\right) ,a,n\right] }\right\} =\quad ;\quad w_o\left(
g,a,n\right) =\ln \left[ x_o\left( g,a,n\right) \right] =  \label{tw}
\end{equation}
where $x_o\left( g,a,n\right) $ is the real and positive solution of the
implicit equation coming from the extremes of the function $G$ in the
asymptotic limit in $\lambda $, namely

\begin{equation}
x_o^2g^2=f\left( x_o,a,n\right) \left[ f\left( x_o,a,n\right) +1\right]
\label{gx}
\end{equation}
and 
\begin{equation}
f\left( x_o,a,n\right) \equiv 1+n+x_o\frac{d\ln \left[ H\left( x_o,a\right)
\right] }{dx_o}  \label{fx}
\end{equation}

In the range of the parameters where $f\left[ x_o\left( g\right) \right] \gg
1$ , we have retained the first order in the expansion of $G$ around the the
saddle point. In so doing, it remains

\begin{equation}
S^{Ap}\left( g,a,n\right) \simeq 4\pi \lambda ^2\left[ 1-\exp \left(
-g/\lambda \right) \right]\, \frac{\exp \left[ G\left( w_o ,t_o ,g,\lambda
,a,n\right) \right] }{\sqrt{D\left( t_o,w_o\right) }}  \label{rep}
\end{equation}
where 
\[
D\left( t_o,w_o\right) \equiv \left. \frac{\partial ^2G}{\partial w^2}%
\right| \left. \frac{\partial ^2G}{\partial t^2}\right| -\left[ \left. \frac{%
\partial ^2G}{\partial wdt}\right| \right] ^2 
\]

Then one checks the positivity condition \cite{jefrey}, in particular when
the discriminant $D\left( t_o,w_o\right) $ of the second derivatives of $G$
at this point is positive. We can now obtain the approximate expression for
the starting amplitude $S\left( g,a,n\right) $ (\ref{int}):

\begin{eqnarray}
S_{Ap}\left( g,a,n\right) &=&\sqrt{2\pi }e^{-1/2}\left[\frac{f\left(
x_o,a,n\right) +1}{D\left( x_o,a,n\right) }\right]^{1/2}\,x_o ^{n+1} 
\nonumber \\
&& H\left( x_o,a\right) \exp \left[ - f\left( x_o,a,n\right) \right]
\label{res}
\end{eqnarray}
where explicitly

\begin{eqnarray}
D\left( x_o,a,n\right) & = & - x_o\,\frac{df\left( x_o,a,n\right) }{dx_o}%
\,\left[\frac{1}{2} + f\left( x_o,a,n\right) \right]  \nonumber \\
& & + f\left( x_o,a,n\right)\,\left[ 1 +f\left( x_o,a,n\right) \right]
\label{dx}
\end{eqnarray}
with $x_o\left( g,a,n\right),and $ $F$ defined above in eq. (\ref{gx}) and (%
\ref{fx}), respectively.

Notice that the expression (\ref{res}) is valid for functions $H\left(
x,a\right) $ that fulfill the following general conditions:

1) the relation (\ref{gx}) must be biunivocal.

2) $D\left( x_o,a,n\right) $ has to be positive and $\left[ x_o\,\frac{
df\left( x_o,a,n\right) }{dx_o}-2\, f\left( x_o,a,n\right) \right] $ has to
be negative in $x_o$ .

3) $f\left( x_o,a,n\right) \gg 1$. This condition is fulfilled in particular
when $n\gg 1.$

These conditions provide the range of values of the involved parameters
where the approximate solution (\ref{res}) is valid. Then, the calculation
of the GBT simply consists on solving the implicit equation (\ref{gx}) to
obtain the saddle point expression. In general, the functional complexity of 
$H$ can add constraints on the range of the parameters.

We are now prepared to study the amplitude coming from the delta potential,
namely the wave packet

\begin{eqnarray}
\Phi \left( Q\right) & \equiv & \frac{\Psi \left( Q\right) }Q=\frac{F\left(
Q\right) +1}Q=\int_0^\infty \frac{\exp \left( -Qx\right) dx}{1+vx/2\left[
1-\exp \left( -2/xm\right) \right] }  \nonumber \\
& & \equiv \int_0^\infty H\left( x,v,m\right) \exp \left( -Qx\right) dx
\label{fi}
\end{eqnarray}

On the other hand, the GBT is able to deal with

\begin{equation}
\Phi _n\left( Q\right) =\int_0^\infty x^nH\left( x,v,m\right) \exp \left(
-Qx\right) dx  \label{finn}
\end{equation}
The relation between (\ref{fi}) and (\ref{finn})

\[
\Phi _n\left( Q\right) =\left( -\right) ^n\frac{\partial ^n}{\partial Q^n}%
\Phi \left( Q\right) 
\]
can be inverted obtaining

\begin{equation}
\Phi \left( Q\right) =\left( -\right) ^n\underbrace{\int dQ\cdots \int dQ}%
_n\Phi _n\left( Q\right) +\sum_{p=0}^{n-1}c_p\left( v,m\right) Q^p
\label{fiapro}
\end{equation}
where the finite sum come from the indefinite integrations. Notice that all
the coefficients vanish whenever the Laplace transform (\ref{fi}) fulfill
the following asymptotic behavior

\[
\lim_{Q\rightarrow \infty }\Phi \left( Q\right) =0 
\]

Moreover, the expression (\ref{fiapro}) is valid for any value of $n,$ in
particular when $n\gg 1.$ Consequently, the approximate solution reads

\begin{equation}
\Phi _{Ap}\left( Q\right) \simeq \left( -\right) ^n\underbrace{\int dQ\cdots
\int dQ}_n\Phi _n^{Ap}\left( Q\right)  \label{fff}
\end{equation}
where, for $n\gg 1$

\begin{eqnarray}
\Phi _n\left( Q\right) &\simeq &\Phi _n^{Ap}\left( Q\right)  \nonumber \\
&=&\sqrt{\frac{2\pi /e\left[ f\left( x_o,n\right) +1\right] }{D\left(
x_o,n\right) }}\,\,\left( x_o\right) ^{n+1}H\left( x_o,v,m\right) \exp
\left[ -f\left( x_o,n\right) \right]  \label{finapro}
\end{eqnarray}
is the approximate solution provided by GBT, with the $D$ and $f$ functions
defined above (\ref{fx}) and (\ref{dx}).

For $n$ sufficiently large and $Q>0$, the explicit expression of the saddle
point becomes $x_o=\left( n+3/2\right) /Q$. Then, we can substitute this
expression into (\ref{finapro}), obtaining the following approximate
solution for the expression (\ref{finn})

\begin{eqnarray*}
\Phi _n^{Ap}\left( Q\right) & \simeq & \frac{\sqrt{2\pi }\left( n+1\right)
^{n/2}\left( 2+n\right) ^{n/2+1/2}\exp \left( -n-3/2\right) }{Q^{n+1}\left\{
1+v\left( n+3/2\right) /2Q\left[ 1-\exp \left( -2Q/\left( n+3/2\right)
m\right) \right] \right\} }  \nonumber \\
& \simeq & \frac{\Gamma \left( n+1\right) }{Q^{n+1}}\frac 1{1+G\left(
n,Q\right) }
\end{eqnarray*}
If $0<Q<\infty $, then $0<G\left( n,Q\right) <v/m<1$ and we can expand to
obtain 
\begin{equation}
\Phi^{Ap} _n\left( Q\right) \simeq \Gamma \left( n+1\right)
\sum_{p=0}^\infty \frac{\left[ -v\left( n+3/2\right) /2\right] ^p}{Q^{p+n+1}}%
\sum_{k=0}^p\binom pk \left( -\right) ^k\exp \left( -2kQ/\left( n+3/2\right)
m\right)  \label{fifin}
\end{equation}

It is illustrative to separate the perturbative and nonperturbative
contributions to this expression. To this end we analyze the first
contribution, e.g., the term $k=0$. This correspond to the asymptotic
solution on $Q$ . After, we add the terms with $k\neq 0$, corresponding to
nonperturbative corrections.

For $k=0$, by solving the $n$-integrations, we obtain

\[
\Phi _o^{GBT}\left( Q\right) \simeq \sum_{p=0}^\infty \frac{\left[ -v\left(
n+3/2\right) /2\right] ^p}{Q^{p+1}}\frac{\Gamma \left( p+1\right) \Gamma
\left( n+1\right) }{\Gamma \left( n+p+1\right) }\simeq \frac
1Q\sum_{p=0}^\infty \frac{\left[ -v/2\right] ^p}{Q^p}\Gamma \left(
p+1\right) 
\]
where in the last step, we have explicitly taken the limit $n\rightarrow
\infty .$ Then, from eq.(\ref{fi})

\[
\Psi _o^{GBT}\left( Q\right) \simeq \sum_{p=0}^\infty \left( -\alpha \left(
Q\right) \right) ^p\Gamma \left( p+1\right) =\Psi _S^{as}\left( Q\right)
=\frac 1{\alpha \left( Q\right) }\exp \left( 1/\alpha \left( Q\right)
\right) \func{Ei}\left( 1,1/\alpha \left( Q\right) \right) 
\]

Consequently, we have recovered the approximate solution (\ref{fias})
provided by the conventional Borel transform in the region where this one
provides sensible results.

To determine the nonperturbative corrections, we must solve the $n-$integrals

\[
I_{n+p+1}\left( \left| a\right| ,Q\right) =\underbrace{\int dQ\cdots \int dQ}%
_n\frac{\exp \left( -\left| a\right| Q\right) }{Q^{n+p+1}}\quad;\quad a\neq
0 
\]
to get 
\[
I_{n+p+1}\left( \left| a\right| ,Q\right) =\frac{\left( -\right)
^{n+p}\left( -\right) ^n}{\Gamma \left( n+p+1\right) \left| a\right| ^{n-1}}%
\frac{d^{n+p}}{db^{n+p}}\left. \exp \left( \left| a\right| b\right) \func{Ei}%
\left( n,\left| a\right| \left( b+Q\right) \right) \right| _{b=0} 
\]
where $\func{Ei}\left( n,x\right) $ is defined in eq. (\ref{expint}). Then,
one can carry out the $\left( n+p\right) $ derivations valued at $b=0$ 
\begin{equation}
I_{n+p+1}\left( \left| a\right| ,Q\right) =\left| a\right| ^{p+1}\exp \left(
-\left| a\right| Q\right) \left( -\right) ^nG\left( n+p+1,p+2,\left|
a\right| Q\right)  \label{iene}
\end{equation}
where $G$ is the confluent hypergeometric functions of second class.

Hence, replacing the expression (\ref{fifin}) into (\ref{fff}) and using eq.(%
\ref{iene}) we can write 
\begin{eqnarray*}
\Psi _S^{GBT}\left( Q\right) &\simeq &\Psi _o^{GBT}\left( Q\right) \\
& &+ \lim_{n\rightarrow \infty }Q\Gamma \left( n+1\right) \sum_{p=1}^\infty
\sum_{k=1}^p\binom pk\left( -\right) ^k\left[ -v\left( n+3/2\right)
/2\right] ^p\left[ \frac{2k}{\left( n+3/2\right) m}\right] ^{p+1} \\
&&\exp \left[ -\frac{2k}{\left( n+3/2\right) m}Q\right] G\left[ n+p+1,p+2,%
\frac{2k}{\left( n+3/2\right) m}Q\right]
\end{eqnarray*}
To perform the limit $n\rightarrow \infty $, we have used here the integral
representation of $G$ . Then, we conclude that the approximate solution reads

\begin{equation}
\Psi _S^{GBT}\left( Q\right) \simeq \Psi _S^{as}\left( Q\right) -4\frac
Qv\sum_{p=1}^\infty \sum_{k=1}^p\binom pk\left( -\right) ^{k+p}\left[ \frac
v{\sqrt{\frac{2m}kQ}}\right] ^{p+1}K_{p+1}\left( 2\sqrt{\frac{2k}mQ}\right)
\label{gbtaprox}
\end{equation}
where $K_p$ is the Bessel function of second class.

The approximate solution (\ref{gbtaprox}) converges rapidly to the exact
solution (\ref{fi}). The nonperturbative corrections recover the
corresponding dependence on $m$ . The deviation with respect to the exact
solution obtained is certainly small within all the range of $Q.$ In
particular, the second order approximation suffer a deviation lower than 4\%
for $Q\gtrsim 2$ (see figure 1).

The test of the accuracy provided by the GBT is presented in Figure 1 where
we have compared the ratio $F_S^{GBT}\left( Q\right) /F\left( Q\right) $
(line $c$) between our approximate analytical solution (\ref{gbtaprox}) up
to second order ( $k=2$ ) and the numerical integration of the exact
expression (\ref{fi}), with the corresponding to the Borel resummation
technique $F_S^{as}\left( Q\right) /F\left( Q\right) $ (line $a$) and
optimization of the PT $F_{\mu =4}\left( Q\right) /F\left( Q\right) $ (line $%
b$). This comparison has been performed for the particular values of the
parameters $m=3$ and $v=0.1.$

Let us now turn to the case of an attractive potential. This corresponds to $%
v<0$, and can be studied by changing $v\rightarrow -\left| v\right| $ in the
previous definition of the potential. In this case the expression for the
wave packet (\ref{sumafi}) is not Borel summable because now its
conventional Borel transform (\ref{sumaborel}) has a pole that does not
allow the integration. This situation is similar to the one presents in
infrared perturbative QCD \cite{altarelli}\cite{weeber}\cite{korch}. In this
case the renormalon technique usually extends the result of Borel summation (%
\ref{fias}) by using the principal value (PV) prescription. Then adds a
term, coming from the ambiguity generated by the pole, that represents the
non perturbative contribution. Consequently, the series (\ref{sumafi})
results in the following approximate solution

\begin{equation}
\Psi _{NS}^{as}\left( Q\right) =\Psi _{PV}^{as}\left( Q\right) +C\exp \left(
1/\alpha \right)  \label{fianp}
\end{equation}
where

\[
\Psi _{PV}^{as}\left( Q\right) =\frac{-1}{\alpha \left( Q\right) }\exp
\left( -1/\alpha \left( Q\right) \right) \overline{\func{Ei}}\left(
1,1/\alpha \left( Q\right) \right) 
\]
and

\begin{equation}
\overline{\func{Ei}}\left( 1,1/\alpha \left( Q\right) \right) \equiv
P.V.\int_1^{-\infty }\frac{\exp \left( t/\alpha \right) dt}t  \label{cauchy}
\end{equation}

The first term in (\ref{fianp}) has the same perturbative asymptotic
expansion as the exact function (\ref{paquete}) and the constant $C$ is
obtained by using purely nonperturbative methods or is extracted from
experiment. The optimal value obtained is $C=-0.06$ (see \cite{rusos}).
However, the results (\ref{fianp}) diverges strongly at small $Q$ (see
figure 2, line $a$).

On the other hand, the alternative approach of the modified perturbation
theory obtain the best approximate results for the value of the parameter $%
\mu =3.6$. The obtained result presents an improvement with respect to the
previous approximate solution reducing the deviation at small $Q$ to
approximately 20\% but its lose the asymptotic exact behavior (see figure 2,
line $b$).

The GBT follows the same steps presented in the case of the repulsive
potential. Notice that previously, the term $k=0$ generates the asymptotic
solution series (\ref{sumafi}) but now this series in non alternate. In this
regime the approximate value for this sum is given by expression (\ref
{cauchy}). The different behavior is related to the non perturbative
contribution. As it was mentioned above, the renormalon technique shows that
this contribution can be estimated through the ambiguity generated by the
non analyticity of the conventional Borel transform. However, as we have
shown in the case of a repulsive potential, the origin is in the remaining
terms with $k\neq 0$ in the expansion of the binomial expression (\ref{fifin}%
). In fact, those terms can not be taking into account in the renormalon
technique. Thus, our approximate solution reads

\[
\Psi _{NS}^{GBT}\left( Q\right) \simeq \Psi _{PV}^{as}\left( Q\right)
-4\frac Q{\left| v\right| }\sum_{p=1}^\infty \sum_{k=1}^p\binom pk\left(
-\right) ^k\left[ \frac{\left| v\right| }{\sqrt{\frac{2m}kQ}}\right]
^{p+1}K_{p+1}\left( 2\sqrt{\frac{2k}mQ}\right) 
\]

As is was the case for the repulsive potential, our approximate solution has
the correct behavior for all the range of the momentum $Q$. In fact, the
deviation, when one considers up to second order ($k=2$), is lower than $%
3.5\%$ for $Q\gtrsim 2$ (see figure 2, line $c$).

In Figure 2 we present the corresponding comparative plots of the ratio
between the approximate solutions $F_{NS}^{as},F_{\mu =3.6},F_{NS}^{GBT}$
and the exact one $F $ for the attractive potential.

In Summary, we have presented an analytic expression for the wave packet
valid for all the range of $Q$, based on the GBT. In the case of the
summable Borel theory, our results show explicitly how the GBT complete the
perturbative solution provided by conventional Borel transform by means of
the adequate incorporation of nonperturbative contributions. In the non
summable Borel theory case, besides obtaining sensible results, we have
shown that the real origin of the nonperturvative contributions is not the
Borel ambiguity.

\pagebreak

\FRAME{ftbpFU}{4.6406in}{3.5907in}{0pt}{\Qcb{numerical comparation for Borel
summable solutions}}{}{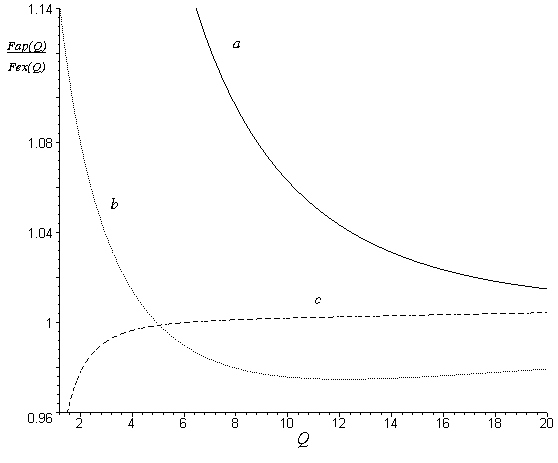}{\special{language "Scientific
Word";type "GRAPHIC";maintain-aspect-ratio TRUE;display "USEDEF";valid_file
"F";width 4.6406in;height 3.5907in;depth 0pt;original-width
419.3125pt;original-height 347.8125pt;cropleft "0";croptop "1";cropright
"1";cropbottom "0";filename 'C:/My
Documents/perturbativo.jpg';file-properties "XNPEU";}} \FRAME{ftbpFU}{%
4.6406in}{3.6002in}{0pt}{\Qcb{numerical comparation for no Borel summable
solutions}}{}{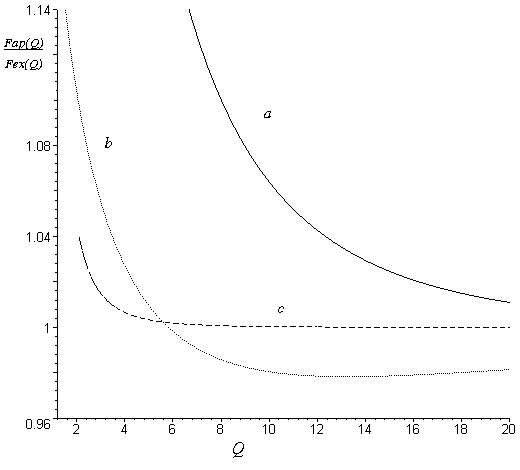}{\special{language "Scientific Word";type
"GRAPHIC";maintain-aspect-ratio TRUE;display "USEDEF";valid_file "F";width
4.6406in;height 3.6002in;depth 0pt;original-width 390.6875pt;original-height
353.0625pt;cropleft "0";croptop "1";cropright "1";cropbottom "0";filename
'C:/My Documents/noperturbativo.jpg';file-properties "XNPEU";}}


\begin{thebibliography}{99}
\bibitem{rusos}  A.A.Penin and A.A.Pivovarov, Phys. Lett. \textbf{B401}, 294
(1997).

\bibitem{hardy}  G.N.Hardy, ''Divergent Series'' , Oxford Univ. Press
(1949); J.C.Le Guillou and J. Zinn-Justin, '' Large Order Behavior of
perturbation Theory'' , North-Holland, Amsterdam (1989).

\bibitem{error}  A.Duncan and S. Pernice, Phys. Rev \textbf{D51}, 1956
(1995); U. Aglietti and Z.Ligeti, Phys. Lett. \textbf{B364}, 75 (1995);
M.Beneke and V.M. Braun, Nuclear Physics \textbf{B426}, 301 (1994).

\bibitem{propuestas}  R. Akhoury and V.I. Zakharov, Phys. Lett. \textbf{B438}%
, 165 (1998); A. V. Nesterenko, Phys. Rev. \textbf{D62}, 094028 (2000); S.J.
Brodsky, P.G.Lepage and P.B. Mackenzie, Phys. Rev. \textbf{D28}, 228 (1983);
P.M. Stevenson, Phys. Rev. \textbf{D23}, 2916 (1981); M. Beneke and V.M.
Braun, Phys. lett. \textbf{B348}, 513 (1995); P. Ball, M. Beneke and V.M.
Braun, Nucl. Phys. \textbf{B452}, 563 (1995).

\bibitem{beneke}  M.Beneke, Phys.Rept. \textbf{317}, 1 (1999).

\bibitem{nos}  L.N.Epele, H. Fanchiotti, C.A. Garc\'{\i }a Canal and M.
Marucho, Nuclear Physics \textbf{B583}, 454 (2000).

\bibitem{nosrich}  L.N.Epele, H. Fanchiotti, C.A. Garc\'{\i }a Canal and M.
Marucho, Phys. Lett. \textbf{B523}, 102 (2001).

\bibitem{dispersion}  W.M.Frank and D.J.Land, Rev.Mod.Physics \textbf{43},
36 (1971).

\bibitem{NU}  A.Nikiforov and V.B.Uvarov, ''Special Functions of
Mathematical Physics'', Cambridge, third edition (1996), pag. 401.

\bibitem{grun}  G.Grunberg, Phys. Lett. \textbf{B327}, 121 (1996).

\bibitem{jefrey}  H.Jeffrey and B.S.Jeffrey, Methods of Mathematical
Physics, Cambridge (1996).

\bibitem{altarelli}  G.Altarelli, Erice Subnuclear, 221 (1995).

\bibitem{weeber}  B.R. Weeber, Phys. Let. \textbf{B399}, 148 (1994); A.V.
Manohar and M.B. Wise, Phys. Lett. \textbf{B344}, 407 (1995).

\bibitem{korch}  G.P. Korchemsky and G. Sterman, Nucl. Phys. \textbf{B437},
415 (1995); M. Neubert and C.T. Sachranjda, Nucl. Phys. \textbf{B438}, 235
(1995).\newpage 
\end{thebibliography}
\end{document}